\documentclass[multphys,vecphys]{svmult}

\usepackage{makeidx}   
\usepackage{graphicx}  
\usepackage{multicol}  

\usepackage{cite}

\makeindex             

\begin{document}

\title{Stochastic Growth in a Small World}

\titlerunning{Stochastic Growth in a Small World}

\author{B. Kozma \and G. Korniss}

\authorrunning{B. Kozma \and G. Korniss}

\institute{
Department of Physics, Applied Physics, and Astronomy, \\
Rensselaer Polytechnic Institute, 110 8th Street,
Troy, NY 12180-3590, USA
\texttt{kozmab@rpi.edu, korniss@rpi.edu}
}

\maketitle

\begin{abstract}
We considered the Edwards-Wilkinson model on a small-world network.
We studied the finite-size behavior of the surface width by performing exact
numerical diagonalization for the underlying coupling matrix. We found that
the spectrum exhibits a gap or a pseudo-gap, which is responsible for
a finite width in the thermodynamic limit for an arbitrarily weak but
nonzero magnitude of the random interactions.
\end{abstract}

\section{Introduction}

Since the introduction of small-world networks
\cite{WATTS,SM_book}, a number of well-known models have been
studied where the original short-range interaction topology is
extended to include a finite number of possibly long-range links
per ``site'' \cite{NEWMAN}. The common observation is that these
systems can undergo a phase transition, even when the random links
are added (or rewired) to a one-dimensional original substrate
\cite{NEWMAN,BARRAT,GITTERMAN,XY_sw,ising_sw,phase_sw}. The nature
of the transition resembles that of a mean-field one
\cite{XY_sw,ising_sw,phase_sw}.

Among the first applications of
small-world networks was to study synchronization in dynamical
systems \cite{SM_book,Strogatz_review} such as the Kuramoto
oscillators \cite{Kuramoto}. The need for autonomous
synchronization for a system with a large number of ``agents''
(processing elements) also
naturally emerges in large-scale parallel discrete-event simulation (PDES)
schemes \cite{FUJI} for systems with short-range interactions and asynchronous
dynamics \cite{LUBA,KORN_PRL}. Frequent and necessary ``local''
communications between processing elements (PEs) to ensure the
asynchronous causal dynamics of the underlying system will eventually
lead to a diverging spread of the progress of the individual PEs
\cite{KORN_PRL,KORN_SCI}. This property can seriously hinder efficient
data collection for such simulation schemes. An alternative to
possibly costly and frequent global synchronizations is to extend the required
short-range communication topology to include ``weak'' random
links \cite{KORN_SCI}. Weak in this context refers to the relative timescale of
actually using the random connections for synchronization.
By directly ``simulating the simulations'' and using simple
coarse-graining arguments, it was demonstrated \cite{KORN_PRL} that the
progress of the PEs with only local synchronization exhibits ``kinetic
roughening'' governed by the Kardar-Parisi-Zhang (KPZ) equation. With
random links added (a finite number per PE) and
invoked at an arbitrarily small (but non-vanishing) rate, however, the
PEs progress in a near-uniform fashion \cite{KORN_SCI}.

Here we focus on how critical fluctuations
(originally present in the steady state of a one-dimensional
system) are suppressed when the interaction topology is extended
to include weak interactions facilitated by random links. To this
end we study the Edwards-Wilkinson (EW) linear stochastic growth
equation on a ``substrate'' with small-world-like topology. This model is
also closely related to phase ordering and synchronization among
coupled oscillators in the presence of noise \cite{phase_sw} and to
the XY-model on a small-world network \cite{XY_sw}. We consider the equation
\begin{equation}
\partial_{t} h_i =
-(2h_i-h_{i+1}-h_{i-1}) - p\sum_{j=1}^{N} J_{ij}(h_i-h_j)
+\eta_{i}(t)\;,
\label{sm_EW}
\end{equation}
where $h_i$ is the surface height, $\eta_{i}(t)$ is a delta-correlated
Gaussian noise with
variance $2$ (without loss of generality), and we have dropped the
$t$-dependence from the argument of $h_i$ for brevity. The matrix $J_{ij}$
represents the (quenched) random links on top of a one-dimensional
lattice of length $N$ (even for simplicity) with periodic boundary
conditions, i.e., $J_{ij}$$=$$1$ if a random link is present and zero
otherwise. The parameter $p$ is the strength of the interaction through the
random links. Our construction of the random links is such that
each site has exactly one random link. More specifically, pairs of sites
are selected at random and once they are chosen,
they cannot be selected again. This somewhat constrained construction of
the random network originates from an application to scalable PDES
synchronization schemes \cite{KORN_SCI}, where fluctuations in the individual
connectivity of the PEs are to be avoided.

For a given realization of this small-world network the average
surface width characterizing the roughness is defined as
\begin{equation}
\langle w^2 \rangle_{N} =
\left\langle\frac{1}{N}\sum_{i=1}^{N}(h_i-\bar{h})^2\right\rangle\;,
\label{w2}
\end{equation}
where $\bar{h}$$=$$(1/N)\sum_{i=1}^{N}h_i$ is the mean height and
$\langle\ldots\rangle$ denotes an ensemble average over the noise
in Eq.~(\ref{sm_EW}). For $p$$=$$0$ in Eq.~(\ref{sm_EW}), we recover
the one-dimensional EW model where the steady-state width diverges as
$\langle w^{2}_{N}\rangle$$=$$N/12$.

One may wonder how the system would behave if the same total
number of links as in the above construction of a small-world
network (i.e., $N/2$) were used to connect each site with the one
located at the ``maximum'' possible distance of away from it
($N/2$ on a ring with periodic conditions). Elementary
calculations show that $\langle w^{2}_{N}\rangle\simeq N/24$ for
large $N$, i.e., the width would diverge as for a one-dimensional
system of size $N/2$. Indeed, one can realize that such regularly
patterned long-range links make the original system equivalent
to a $2$$\times$$(N/2)$ system with {\em only} nearest-neighbor
interactions and shifted periodic boundary conditions. More
generally, one can show that, if every site is connected to a
finite number of others and the length of those links can only
assume a {\em finite} set of long-range values (all scaling with
$N$), the width will still diverge in the same fashion as for the
one-dimensional case.

\section{Roughness and the Density of States}

We study the finite-size effects of the width of the surface and also the
underlying spectrum (density of states) of the associated random
matrix which governs the steady-state height fluctuations.
Exploiting that the noise in Eq.~(\ref{w2}) is Gaussian, the
steady-state width for a single realization of the random network can
be expressed as
\begin{equation}
\langle w^2 \rangle_{N} = \frac{1}{N}\sum_{k=1}^{N-1}\frac{1}{\lambda_k}
\label{w2_lambda}
\end{equation}
where $\lambda_k$ are the eigenvalues of the real symmetric coupling matrix
\begin{equation}
\Gamma_{ij} =
\left\{(2+p)\delta_{ij}-\delta_{i+1\, j}-\delta_{i-1\, j}\right\} - pJ_{ij}
\label{gamma}
\end{equation}
as can be read off from Eq.~(\ref{sm_EW}). Note that we have
exploited our specific choice of $J_{ij}$, resulting in
$\sum_{j=1}^{N}J_{ij}$$=$$1$ for all $i$. Also, note that since
Eq.~(\ref{w2}) contains the height fluctuations measured from the
mean, the eigenvalue $\lambda_0$$=$$0$, corresponding to the uniform
eigenvector (zero-mode) of $\Gamma_{ij}$, does not appear in
Eq.~(\ref{w2_lambda}). In the limit of $N$$\to$$\infty$ and assuming
that the distribution of the eigenvalues of $\Gamma_{ij}$ becomes
self-averaging, the disorder-averaged width can be be written as
\begin{equation}
\left[\langle w^2 \rangle_{N}\right] = 
\left[\frac{1}{N} \sum_{k=1}^{N-1}\frac{1}{\lambda_k}\right] 
\stackrel{N\to\infty}{\longrightarrow}
\int\!\frac{\rho(\lambda)d\lambda}{\lambda} \;,
\label{dos}
\end{equation}
where $[\ldots]$ stands for averaging over the
random-link disorder and $\rho(\lambda)$ denotes the density of
eigenvalues of $\Gamma_{ij}$. The behavior of $\rho(\lambda)$ as
$\lambda$ goes to zero determines whether the width remains finite or
diverges in the thermodynamic limit. In the pure one-dimensional case,
$\rho(\lambda)$ actually diverges as $1/(2\pi\sqrt\lambda)$. If, however,
$\rho(\lambda)$ exhibits a gap or approaches zero fast enough,
$\left[\langle w^2 \rangle_{N}\right]$ will be finite.
In the context of diffusion on a
small-world network, it was found that the density of states exhibits a
pseudo-gap (vanishes exponentially fast) \cite{MONA}. The construction of
the small-world graph in Ref.~\cite{MONA}  allowed for the existence of
arbitrarily long ``pure'' chain-segments of the network with exponentially
small probabilities. These small, but non-vanishing, probabilities were
responsible for the pseudo-gap \cite{BRAY}. In our specific
construction of the network, where each site has exactly one random link,
the above argument does not apply and a true gap may develop. Further,
the coupling matrix [Eq.~(\ref{gamma})] has a realization-independent ``mass''
term. This property would actually allow for a perturbation expansion
for small but non-zero values of $p$ with the term $-pJ_{ij}$ being the
perturbation. 

We performed exact numerical diagonalization of the coupling matrix
Eq.~(\ref{gamma}) using standard numerical routines \cite{NUM_REC},
and calculated the
steady-state width as a function of the system size for various values of
$p$. The results are summarized in Fig.~1(a). We also plotted the
analytic form of the width for the simple ``massive'' coupling
matrix, the expression in brackets in Eq.~(\ref{gamma}),
as the zeroth-order approximation in a perturbative
approach. It appears that for small values of $p$, the numerically
computed (and disorder-averaged) width and this simplest
approximation yield the same asymptotic finite-size effects.
Fig.~1(b) shows the cumulative eigenvalue distribution
$\int^{\lambda}\!\rho(\lambda^{\prime})d\lambda^{\prime}$ for $N$$=$$1000$
for various $p$ values. Whether the spectrum exhibits a true or a
pseudo-gap (due to exponentially small likely eigenvalues), cannot
be determined by numerics. It is appears, however, that the
numerically observed ``gap'' asymptotically scales linearly with small
values of $p$. 
\begin{figure}[t]
\centering
\includegraphics[width=.48\textwidth,height=4.7truecm]{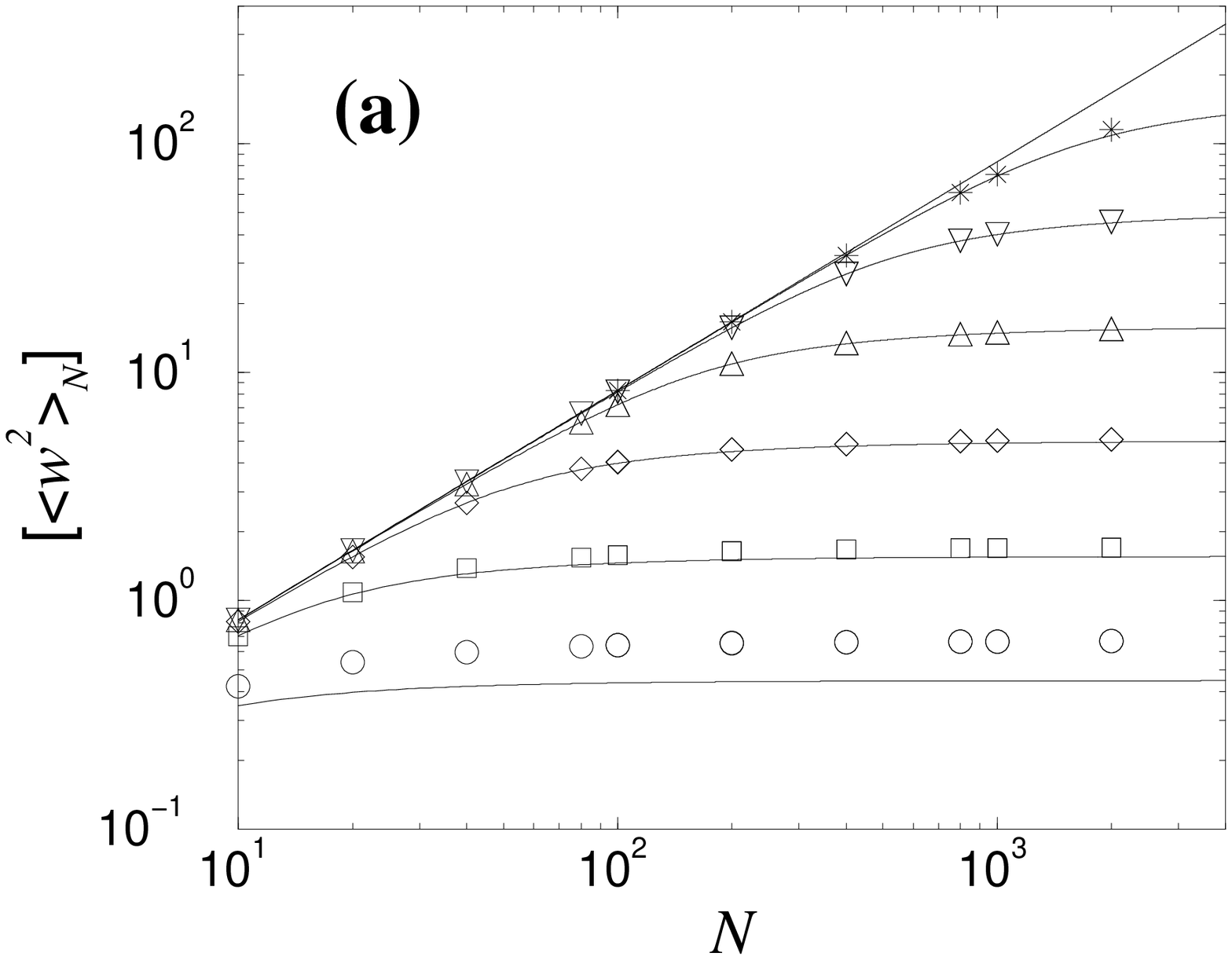}
\hspace*{0.2truecm}
\includegraphics[width=.48\textwidth,height=4.7truecm]{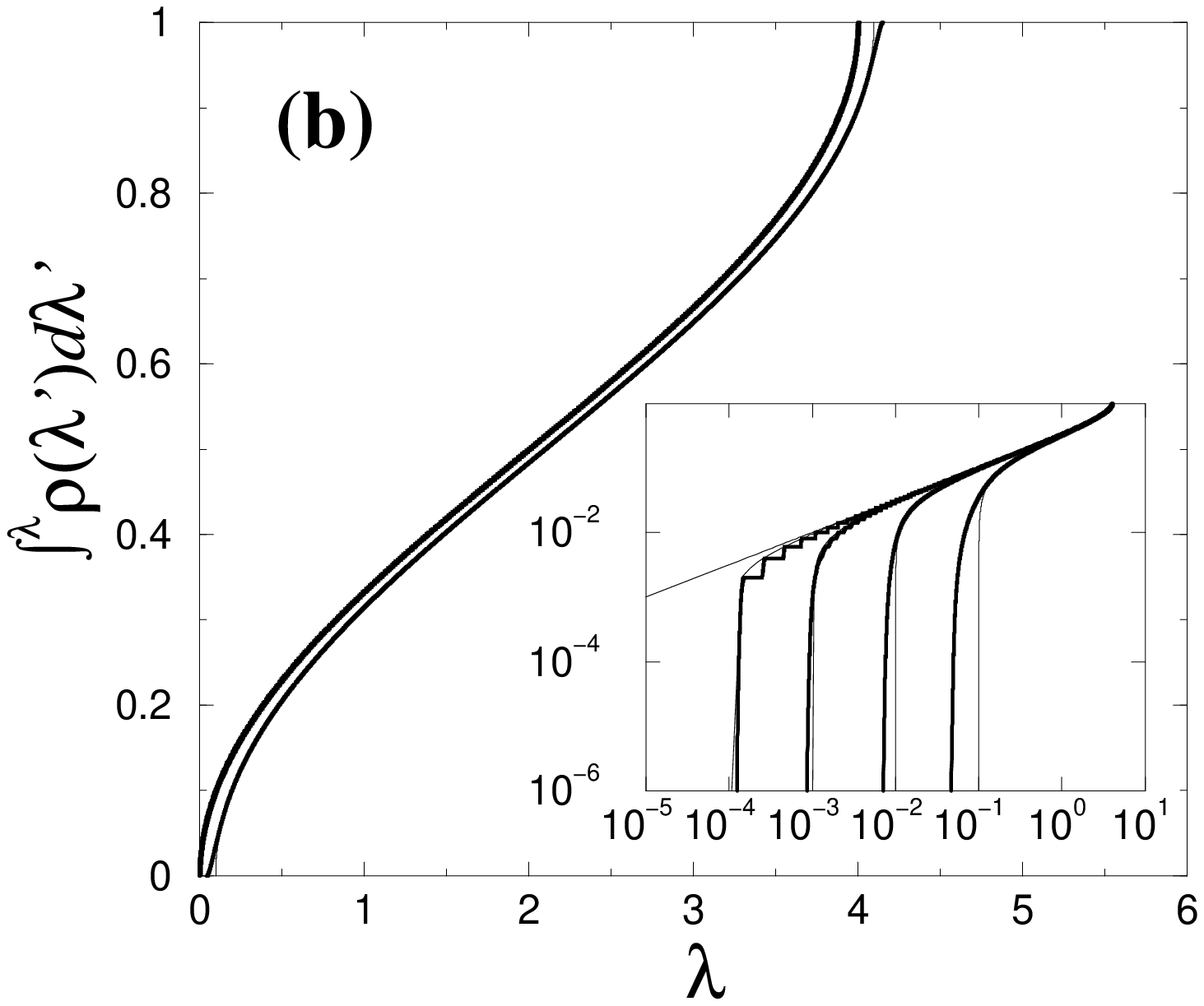}
\caption[]{
({\bf a}) Disorder-averaged surface width
$[\langle w^{2}_{N} \rangle]$ as a function of the system size $N$ for
$p=1,10^{-1},10^{-2},10^{-3},10^{-4}$, and $10^{-5}$ (from bottom to top,
respectively, with symbols), obtained by using $1000$ independent
realizations of 
the random network (except for $N$$=$$2000$ where only $100$ realizations were
generated). The solid lines are the analytic forms for the
width in the simple ``massive'' approximation. The straight line
corresponds to the ``rough'' $p$$=$$0$ case.
({\bf b}) Cumulative (integrated) density of states for
$p=10^{-1},10^{-2},10^{-3}$, and $10^{-4}$ (from right to left, respectively,
with bold solid lines) based on $1000$ realizations of the random
network for $N$$=$$1000$. The thin lines correspond to the analytic
form for the infinite system-size ``massive'' approximation. These
curves are not distinguishable on normal scales, except for the $p$$=$$10^{-1}$
case. The inset shows the 
same on log-log scales to magnify the region near small eigenvalues. We
also plot the analytic form for the infinite-system $p$$=$$0$ case
(asymptotically a straight line [$\sim (1/\pi)\sqrt{\lambda})$] for small
$\lambda$.} 
\label{fig1}
\end{figure}

\section{Conclusions}
We carried out exact numerical diagonalization for the coupling
matrix representing EW growth on a small-world network. In our
construction each site had one random link, i.e., no fluctuations
were allowed in the connectivity. We found that the surface width
saturates for all nonzero values of the amplitude of the random
coupling as a result of the gap or pseudo-gap in the underlying spectrum. 
We should also note the similarity between the
relaxation properties of our model and that of the one-dimensional
Ising-like systems with ("annealed") random spin-exchange process
\cite{DROZ} . This long-range process creates a mean-field-like
environment, in which ordering is possible with other suitable
chosen {\em local} processes present \cite{DROZ}.

\section*{Acknowledgments}

Discussions with Z. R\'acz, G. Gy\"orgyi, P.A. Rikvold, M.A.
Novotny and Z. Toroczkai and their careful reading of the
manuscript are gratefully acknowledged. This research is supported
in part by US NSF through Grant No.\ DMR-0113049 and the Research
Corporation through Grant No.\ RI0761.

%

\end{document}